\documentclass[aps,prd,10pt,a4paper,altaffilletter,amssymb,showpacs,nofootinbib,twocolumn,showkeys,a4paper,floatfix]{revtex4-1}
\usepackage{graphicx,epsfig}
\usepackage{subfig}
\usepackage{amsmath}

\begin{document}

\title{Observational constraints from SNe Ia and Gamma-Ray Bursts on a clumpy universe}
\author{Nora Bret\'on$^{}$ and Ariadna Montiel$^{}$.}

\affiliation{$^{}$ Departamento de F\'{\i}sica, Centro de Investigaci\'on y de Estudios Avanzados del I. P. N., Apartado Postal 14-740, 07000 M\'exico D.F., Mexico.}

\begin{abstract}
The luminosity distance describing the effect of local inhomogeneities in the propagation of light proposed by Zeldovich-Kantowski-Dyer-Roeder (ZKDR) is tested with two probes for two distinct ranges of redshifts: supernovae Ia 
(SNe Ia) in $0.015 \le  z \le 1.414$ and gamma-ray bursts (GRBs) in $1.547 \le z \le 3.57$.
Our analysis is performed by a Markov Chain Monte Carlo (MCMC) code that allows us to constrain the matter density parameter $\Omega_m$ as well as the smoothness parameter $\alpha$ that measures the inhomogeneous-homogeneous rate of the cosmic fluid in a flat $\Lambda$CDM model.
The obtained best fits are $(\Omega_m=0.285^{+0.019}_{-0.018}, \alpha= 0.856^{+0.106}_{-0.176})$ from SNe Ia and
$(\Omega_m=0.259^{+0.028}_{-0.028}, \alpha=0.587^{+0.201}_{-0.202})$ from GRBs, while from the joint analysis 
the best fits are $(\Omega_m=0.284^{+0.021}_{-0.020}, \alpha= 0.685^{+0.164}_{-0.171})$ with a $\chi^2_{\rm red}=0.975$. The value of the smoothness parameter $\alpha$ indicates a clumped universe however it does not have an impact on the amount of dark energy (cosmological constant) needed to fit observations. This result may be an indication that the Dyer-Roeder approximation does not describe in a precise form the effects of clumpiness in the expansion of the universe.
\end{abstract}


\maketitle

\section{Introduction}
One of the most important problems in modern cosmology is the coincidence in the order of magnitude between the inferred energy density of the source driving the acceleration and the energy density of matter. Since the most evident qualitative change in the late universe is the formation of nonlinear structures, one cannot help noticing that cosmic acceleration and the formation of large structure occur, coincidently, in near epochs. If the recent formation of large structures could be identified as the driver for cosmic acceleration, the coincidence would be explained in a natural way. In fact there are theoretical results that prove that gravitational collapse can lead acceleration, see for example \cite{Rasanen2006}. Moreover, in Refs. \cite{Buchert2011} and \cite{Kolb2011}, it is argued that the influence of inhomogeneities on the effective evolution history of the universe is encoded in backreaction terms, and that these terms can mimic dark energy;
in this context see also Ref. \cite{Clarkson2011} for a review on different approaches to the dynamical aspects of backreaction effects in recent eras.

It has also been addressed by several authors (see for instance Refs. \cite{Wiltshire2007,Ellis2009,Clarkson2011}, that the process of averaging and solving the Einstein equations is non-commutative and that maybe if we do it carefully, the accelerated expansion lately observed can be explained without the need of dark energy.

On the other hand, most cosmological observations probe quantities related to light propagation such as the redshift, the angular diameter distance, the luminosity distance or image distortion. The prevailing model $\Lambda$CDM assumes that this propagation occurs through a homogeneous perfect fluid; but on scales smaller than 100 $h^{-1}$Mpc it is not reliable to consider that the universe can be modeled by a FRW cosmology. 

The incorporation to the FRW model of local inhomogeneities have been addressed by several authors since the sixties, namely, works by Zel'dovich \cite{Zeldovich}, Bertotti \cite{Bertotti}, Kantowski \cite{Kantowski1969} and, more recently, by Dyer and Roeder \cite{Dyer}, again by Kantowski \cite{Kantowski1998} and by Demianski \cite{Demianski}.
The general idea is to consider the propagation of light rays in a fluid that may clump. The proportion of clumped matter respect to the homogeneous fluid is measured by the smoothness parameter. Departing from the Sachs expression for a congruence of null geodesics, neglecting twist and shear effects, and including in the matter content the smoothness parameter, they arrive to an equation for the angular diameter distance, that can be connected, via the Etherington relation, to the observable luminosity distance, resulting then the so called Zeldovich-Kantowski-Dyer-Roeder (ZKDR) luminosity distance.  

In this paper we consider observations of SNe Ia in the range of $0.015 \le z \le1.414$ \cite{Union21} and GRBs in the range of $1.547 \le z \le 3.57$ \cite{Yonetoku12} to test the ZKDR distance and constrain the smoothness or clumpiness  parameter. For our analysis we employ the Markov Chain Monte Carlo (MCMC) technique, it leads us to obtain better constraints than previous analysis \cite{Busti2}.

Furthermore, the GRBs sample is a very improved one with a significative diminished dispersion in the data \cite{Yonetoku12}. From the joint analysis, SNe Ia and GRBs, the best fits are $(\Omega_m=0.284^{+0.021}_{-0.020}, \alpha=0.685^{+0.164}_{-0.171})$ with a $\chi^2_{\rm red}=0.975$, indicating a clumped universe. Despite this fact, the amount of the dark energy (cosmological constant) needed to fit observations is not less than in the homogeneous FRW model. This may be an indication that the Dyer-Roeder approach is missing something in its physical description and should be revised, see Ref. \cite{Clarkson2012} and references therein for a further discussion.
 
In the literature the ZKDR luminosity distance has been probed also with Hubble parameter measurements. Similarly, we include this test using the data reported in Ref. \cite{Jimenez12}, obtaining from the joint analysis (SNe Ia, GRBs and Hubble)
the best fits:  $\Omega_m=0.275^{+0.019}_{-0.018}$ and $\alpha=0.821^{+0.110}_{-0.129}$ with a $\chi_{\rm red}^2=0.974$ that improves the previous analysis reported in Ref. \cite{Busti1}.

The paper is organized as follows: in Sec. II we briefly sketch how to obtain the ZKDR luminosity distance from the null geodesic congruence in a clumped universe, using an exact solution of a hypergeometric equation; in Sec. III  the observational data and statistical method are described. In Sec. IV we discuss the results and finally
in the last Section concluding remarks are given.

\section{The ZKDR luminosity-distance}

Let us consider the light propagation in the geometric optics approximation. The two optical scalars, expansion and shear, $\theta$ and $\sigma$, respectively,  satisfy the Sachs propagation equations \cite{Sachs},
\begin{equation}
\dot{\theta}+\theta^2+{\sigma^2}=-\frac{1}{2}R_{\alpha\beta}k^{\alpha} k^{\beta},
\label{theta}
\end{equation}

\begin{equation}
\dot{\sigma}+2\theta \sigma=-\frac{1}{2}C_{\alpha\beta\gamma\delta} \overline{m}^{\alpha}
k^{\beta} \overline{m}^{\gamma} k^{\delta},
\end{equation}
where a dot denotes differentiation with respect to the affine parameter $\lambda$ and $k^{\mu}$ is the null vector field tangent to the light ray. Notice that since we shall address the RW geometry, the Weyl tensor is zero and the shear vanishes. In \cite{Sasaki1993} it is shown that neglecting the Weyl focusing seems to be correct in our universe.

The expansion $\theta$ is related to the relative change of an infinitesimal surface area A of the beam's cross section by
\begin{equation}
\theta=\frac{1}{2}\frac{d \ln A}{d\lambda}= \frac{1}{2} \frac{\dot{A}}{A}.
\label{theta_A}
\end{equation}

Since the angular diameter distance $D_{A}$ is proportional to $\sqrt{A}$, using $D_{A}$ instead of $\sqrt{A}$ and
considering the affine parameter $\lambda$ as a function of the redshift $z$, it is obtained 
\begin{equation}
\left(\frac{dz}{d\lambda}\right)^2\frac{d^2D_{A}}{dz^2}+
\left(\frac{d^2z}{d\lambda^2}\right)\frac{dD_{A}}{dz}+\frac{4\pi
G}{c^4}T_{\alpha\beta}k^\alpha k^\beta D_{A}=0,
\label{add-r}
\end{equation}
where we have used Einstein's equations to replace $R_{\alpha\beta}$ in terms of $T_{\alpha\beta}$.

A solution of the last expression is related to the angular diameter distance if it satisfies the initial conditions that make the wavefront satisfy Euclidean geometry when leaving the source:
\begin{equation}
D_{A}(z)\mid_{z=0}=0, \quad \frac{dD_{A}(z)}{dz}\mid_{z=0}=\frac{c}{H_0},
\end{equation}
where $H_0$ is the present value of the Hubble parameter and $c$ is the velocity of light.

In order to take into account local inhomogeneities in the distribution of matter, Dyer and Roeder \cite{Dyer} considered that the matter density $\rho_m$ should be interpreted as the average density of the intergalactic regions where the observed light has traveled. Then to measure this effect they introduced a phenomenological parameter $\alpha =(1-{\rho_{\rm clumps}}/{\rho})$ called the clumpiness or smoothness parameter, which quantifies the amount of matter in clumps relative to the amount of matter uniformly distributed. Its maximum value  $\alpha=1$ corresponds to the homogeneous FRW case, while for a totally clumped universe $\alpha=0$; an intermediate value $0 < \alpha <1$ indicating a partially clumped universe.

Assuming an energy-momentum tensor of the form $T_{\mu \nu}= \alpha \rho_m u_{\mu}u_{\nu}+ \rho_{\Lambda} g_{\mu \nu}$, where $u_{\mu}$ is the four-velocity of a comoving volume element, $\rho_m$ is the matter energy density and $\rho_{\Lambda}={\Lambda}/8 \pi G$ is the vacuum energy density associated to the cosmological constant, Eq. (\ref{add-r}) can be rewritten in the form
\begin{equation}
\left(\frac{dz}{d\lambda}\right)^2\frac{d^2D_{A}}{dz^2}+\left(\frac{d^2z}{d\lambda^2}\right)\frac{dD_{A}}{dz}+\frac{3}{2}\alpha \Omega_m(1+z)^5D_{A}=0.
\label{add}
\end{equation} 

We shall consider a flat (zero curvature), homogeneous and isotropic geometry of Robertson-Walker (RW), including only dark matter (dust) and dark energy components. The corresponding Hubble parameter is given by
\begin{equation}
H(z)= \frac{\dot a}{a}=H_0\sqrt{\Omega_m(1+z)^3+\Omega_\Lambda}.
\label{H(z)}
\end{equation}

On the other hand, transforming the affine parameter to the observable redshift by using
\begin{equation}
\frac{dz}{d\lambda}=(1+z)^2\frac{H(z)}{H_0},
\label{z-lambda}
\end{equation} 
and an appropriate change of variable,
\begin{equation}
h(D_A,z)=(1+z)D_A, \quad \zeta(z)= \frac{1+ \Omega_m z(3+3z+z^2)}{1- \Omega_m},
\label{arg}
\end{equation}
Eq. (\ref{add}) can be transformed into a hypergeometric equation
\begin{equation}
(1-\zeta)\zeta \frac{d^2 h}{d \zeta^2}+ \left({\frac{1}{2}- \frac{7}{6}\zeta}\right) 
\frac{d h}{d \zeta}+ \frac{\nu (\nu+1)}{36}h=0,
\label{hyperg_eq}
\end{equation}
where the parameter $\nu$ is related to the clumpiness parameter, $\alpha$, by 
\begin{equation}
\alpha=\frac{1}{6}(3+\nu)(2-\nu).
\label{alpha-nu}
\end{equation}
  
The range for $\nu$ is $0 \le \nu  \le 2$, where $\nu=0 (\alpha=1)$ corresponds to a FRW fluid, while $\nu=2 (\alpha=0)$ to a totally clumped universe.

Since $(1- \Omega_m)<1$, the variable in Eq. (\ref{hyperg_eq}) satisfies $\zeta >1$; thus to make it compatible with the defined range for the argument of the hypergeometric functions, we transform to the inverse argument, $\zeta^{-1} <1$, using properties of the hypergeometric functions.

To connect the angular diameter distance $D_A$ with the observable luminosity distance $d_L$, we use the Etherington relation \cite{Etherington}, $d_L=D_{A}(1+z)^2$. Incorporating also the initial conditions, the luminosity distance we used for the observational tests is given by \cite{Kantowski2001},

\begin{widetext}
\begin{eqnarray}
d_L(\Omega_m, \nu; z) & = & \frac{c}{H_0} \frac{2(1+z)}{\Omega_m^{1/3}(1+2 \nu)}  [1+ \Omega_m z(3+3z+z^2)]^{\nu/6} \nonumber\\
&& \times \left\{  {}_2 F_1\left( -\frac{\nu}{6}, \frac{3-\nu}{6}; \frac{5-2\nu}{6};\frac{1-\Omega_m}{1+\Omega_m z(3+3z+z^2)} \right) \right. \nonumber\\ 
&& \times {}_2F_1\left(\frac{1+\nu}{6}, \frac{4+\nu}{6};\frac{7+2\nu}{6};1-\Omega_m \right) \nonumber\\
&& -[1+ \Omega_m z(3+3z+z^2)]^{-(1+2\nu)/6} {}_2F_1\left(-\frac{\nu}{6}, \frac{3-\nu}{6};\frac{5-2\nu}{6};1-\Omega_m \right) \nonumber\\
&& \left. \times {}_2F_1\left( \frac{1+\nu}{6}, \frac{4+\nu}{6};\frac{7+2\nu}{6};\frac{1-\Omega_m}{1+ \Omega_m z(3+3z+z^2)}   \right) \right\}.
\label{D_L}
\end{eqnarray}
\end{widetext}
 
\begin{figure}
\includegraphics[width=0.45\textwidth]{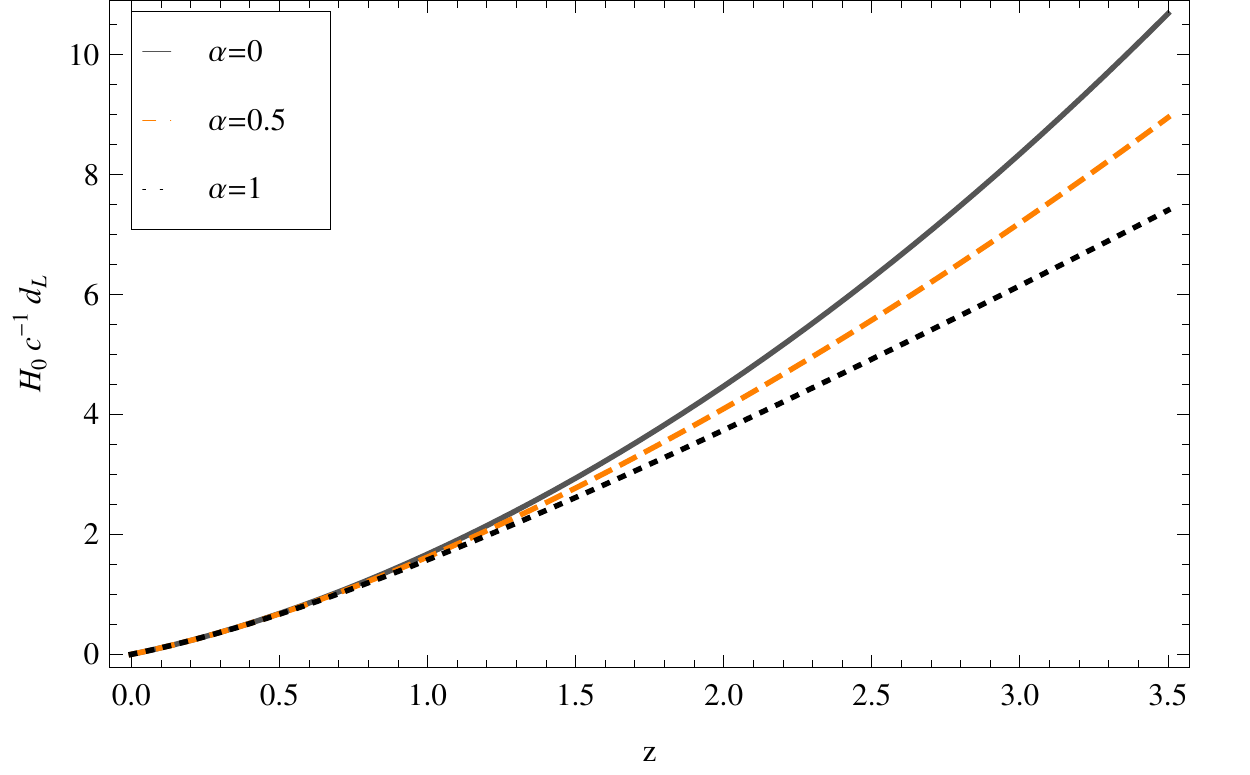}
\caption{The plot corresponds to the ZKDR luminosity distance as a function of the redshift $z$, from Eq. (\ref{D_L}); we have fixed $\Omega_m=0.266 \pm 0.029$ from WMAP-7 years \cite{WMAP} and plot for different values of the smoothness parameter: $\alpha=0$ (a completely clumped universe), $\alpha=1$ (homogeneous FRW) and for a partially clumped universe, $\alpha=0.5$. Clearly the effect of diminishing the smoothness parameter is to increase the luminosity distance.}
\label{d_L(z)}
\end{figure}

A few comments are in order: In Eq. (\ref{theta}), for the rate of expansion of the null congruence, the Ricci tensor, $R_{\alpha \beta}$, makes null rays converge. Then it is reasonable to expect that diminishing $R_{\alpha \beta}$ would result into reducing the convergence of the null bundle, i.e. there would be a dimming of light rays.
This effect can be done by diminishing the density of matter $\rho_m$ through the introduction of the smoothness parameter, $\rho_m \mapsto \alpha \rho_m$, $0 \leq \alpha \leq 1$.
The plots in Fig. \ref{d_L(z)} for the ZKDR luminosity distance versus the redshift for different values of the smoothness parameter, show that this actually happens so:
as $\alpha$ is smaller, $d_L$ grows faster.
 

Have in mind that it is a naive reasoning, because neither Eq.(\ref{theta}) nor  Eq.(\ref{theta_A}) are linear equations.  
Notice that in this approach the luminosity distance, Eq. (\ref{D_L}), does not totally capture the effect of inhomogeneities in the expansion rate, since in the derivation of $d_L$ are assumed the same expressions for $H(z)$ and $\lambda(z)$  than in the homogeneous FRW model, Eqs. (\ref{H(z)})-(\ref{z-lambda}). However, the effects of local inhomogeneities seem to be more complex to be described only through the smoothness parameter.

In this regard, in Ref. \cite{Clarkson2012} it is quantified the density probability distribution as a function of the beamwidth of supernovae and it is discussed a modified version of the DR approximation. It is shown that even for Gpc-lenght beams of 500 kpc diameter,  nonlinear corrections appear to be non-trivial, being not even clear whether underdense regions lead to dimming or brightening of sources (see Ref. \cite{Bolejko2013}).
It was already noted by Linder in Ref. \cite{Linder1988} that the smoothness parameter is a function of the redshift.
This dependence is also considerer in Ref. \cite{Mattsson}, where a generalized Dyer-Roeder distance is introduced. However as was recognized in Ref. \cite{Bolejko2011}, this should be considered carefully because if the smoothness parameter changes with the redshift, the luminosity distance expression will depend on several free parameters with difficult physical interpretation. A more natural dependence of the smoothness parameter with the redshift, coming from weak lensing, is presented in Ref. \cite{Bolejko2011}. In Ref. \cite{Sereno} it was addressed
the properties of cosmological distances in inhomogeneous quintessence cosmology. 

Although the DR approach has been criticized by several authors, the contrast of $d_L$ versus observational data so far has not lead to conclusive results, in the sense of totally discarding this model.
Therefore it is important to elucidate its validity and scope in the description of the universe. So for the time being, we look for useful information from the simplest approach that considers $\alpha$ being constant.

\section{Testing the ZKDR distance vs. observations}

\subsection{Samples and statistical method}

We have used the  updated observational data from Union2.1 supernovae data set reported in Ref. \cite{Union21} and the distance modulus of long Gamma-ray Bursts reported in Ref. \cite{Yonetoku12} to constrain the two free parameters of the model, the matter density $\Omega_m$, and the smoothness parameter  $\nu$ (related to $\alpha$ through Eq. (\ref{alpha-nu})). In order to do that, we used a Markov Chain Monte Carlo (MCMC) code to maximize the likelihood function $\mathcal{L}(\theta_i) \propto \exp [-\chi^2(\theta_i)/2]$ where $\theta_i$ is the set of model parameters and the expression for $\chi^2(\theta_i)$ depends on the dataset used. The MCMC methods (completely described in Refs. \cite{Berg,MacKay,Neal} and references therein) are well-established techniques for constraining parameters from observational data. To test their convergence, here we follow the method developed and fully described in Ref. \cite{Dunkley05}. Additionally, we have tested the ZKDR luminosity distance with the observational Hubble parameter data given in Ref. \cite{Jimenez12}.

\subsection{Supernovae Type Ia (SNe Ia)}
Hitherto we know one of the cosmological observations highly capable to measure directly the expansion rate of the universe are the Supernovae Type Ia. In this work, we use the updated compilation released by the Supernova Cosmology Project (SCP): the Union2.1 compilation which consists of 580 SNe Ia \cite{Union21}. The Union2.1 compilation is the largest published and spectroscopically confirmed SNe Ia sample to date. Constraints from the SNe Ia data can be obtained by fitting the distance moduli $\mu(z)$. The distance modulus can be calculated as
\begin{eqnarray}
\mu(z_j)&=& 5 \log_{10} [D_L(z_j, \theta_i) ] + 25 \nonumber \\
        &=& 5 \log_{10} [d_L(z_j, \theta_i) ] + \mu_0,
\label{Eq:mu}
\end{eqnarray}
where $\mu_0=42.38-5\log_{10} h$, $h$ is the Hubble constant $H_0$ in units of 100 km s$^{-1}$Mpc$^{-1}$, and $d_L(z_j,\theta_i)$ is the ZKDR luminosity distance given by Eq. (\ref{D_L}); $\theta_i$ denotes the vector model parameters $(\Omega_m, \nu)$. The $\chi^2$ function for the SNe Ia data is given by
\begin{equation}
\chi^2_\mu (\mu_0, \theta_i)= \sum^{580}_{j=1} \frac{[\mu(z_j; \mu_0, \theta_i)-\mu_{obs}(z_j)]^2}{\sigma^2_{\mu}(z_j)},
\end{equation}
where the $\sigma^2_{\mu}$ corresponds to the error on distance modulus for each supernova. The parameter $\mu_0$ in Eq. (\ref{Eq:mu}) is a nuisance parameter since it encodes the Hubble parameter and the absolute magnitude $M$, so it is more convenient to marginalize over it. 

Here, we maximize the likelihood by minimizing $\chi^2$ with respect to $\mu_0$, as suggested in \cite{Pietro03,Nesseris05}.  Then one can rewrite the $\chi^2$ function as
\begin{equation}
\chi^2_{SN} (\theta)= c_1 - 2c_2 \mu_0 + c_3 \mu^2_0,
\end{equation}
where
\begin{equation}
c_1=\sum^{580}_{j=1} \frac{[\mu(z_j; \mu_0=0,\theta_i)-\mu_{obs}(z_j)]^2}{\sigma^2_{\mu}(z_j)},
\end{equation}

\begin{equation}
c_2=\sum^{580}_{j=1} \frac{\mu(z_j; \mu_0=0,\theta_i)-\mu_{obs}(z_j)}{\sigma^2_{\mu}(z_j)},
\end{equation}

\begin{equation}
c_3=\sum^{580}_{j=1} \frac{1}{\sigma^2_{\mu}(z_j)}.
\end{equation}

The minimization over $\mu_0$ gives $\mu_0=c_2/c_3$. So the $\chi^2$ function takes the form
\begin{equation}
\tilde{\chi}_{SN}^2 (\theta_i)= c_1 - \frac{c^2_2}{c_3}.
\end{equation}

Since $\tilde{\chi}^2_{SN}=\chi^2_{SN}(\mu_0=0,\theta_i)$ (up to a constant), we can instead minimize $\tilde{\chi}^2_{SN}$ which is independent of $\mu_0$.

\subsection{Long gamma-ray bursts (LGRBs)}

Previous analysis have shown that the ZKDR distance at short distances (redshifts in the range $0.1 \le z \le 1.7$) does not make a difference respect to the $\Lambda$CDM model; it is then appealing to test the model for larger distances, like the ones of the GRBs, that extend the redshift range as far as 8.1. However, the problem is that GRBs appear not to be standard candles and to extract cosmological information it is necessary calibrate them for each cosmological model tested.

There have been several efforts to calibrate the correlations between the luminosity and spectral properties of GRBs in a cosmology-independent way and some proposals to use SNe Ia measurements to calibrate them externally are given in Refs. \cite{Kodama08,Liang08,Wei09}.

More recently, in Ref. \cite{Yonetoku12}, it has been estimated the distance modulus to long gamma-ray bursts (LGRBs) using the Type I Fundamental Plane, a correlation between the spectral peak energy $E_p$, the peak luminosity $L_p$, and the luminosity time $T_L\equiv E_{\rm iso}/L_p$, where $E_{\rm iso}$ is the isotropic energy. Basically the calibration was done in this way: first, the Type I Fundamental Plane of LGRBs was calibrated using 8 LGRBs with redshift $z<1.4$ and SNe Ia (Union2) in the same redshift range by a local regression method, to avoid any assumption on a cosmological model; then this calibrated Type I Fundamental Plane was used to measure the distance modulus to 9 high-redshift LGRBs (see Ref. \cite{Yonetoku12} for calibration's details).  We used the 9 calibrated LGRBs reported in Table~2 of Ref. \cite{Yonetoku12} to derive constraints on $\Omega_m$ and $\nu$ from the ZKDR luminosity distance.
The $\chi^2$ function for the GRBs data is defined by
\begin{equation}
\chi^2(\theta_i )= \sum_{j=1}^{9} \left[\frac{\mu(z_j, \theta_i) - \mu_{\rm obs}(z_j)}{\sigma_{\mu_j}}\right]^{2},
\end{equation}
where $\mu(z_j)= 5 \log_{10} [d_L(z_j, \theta_i) ] + 25$. Notice that we have used the standard expression of $\chi^2$ given through the observed distance moduli just to be consistent with the way in which the calibration was done. We have fixed $H_0$ as $73.8\pm2.4$ from Ref. \cite{Riess}.

\subsection{Hubble parameter observations}

We use the compilation of Hubble parameter measurements estimated with the differential evolution of passively evolving early-type galaxies as cosmic chronometers, in the redshift range $0.09 \le z \le 1.75$ recently updated in Ref. \cite{Jimenez12} but first reported in Ref. \cite{Jimenez02}.

This approach consists in the measurement of the differential age evolution of these chronometers as a function of redshift, in this way a direct estimate of the Hubble parameter $H(z) = -1/(1 + z)dz/dt \simeq -1/(1 + z)\Delta z/\Delta t$ is obtained, providing a reliable differential quantity $\Delta z/ \Delta t$, with many advantages in minimizing common issues and systematic effects. Compared with other techniques, it provides a direct measurement of the Hubble parameter, and not of its integral, in contrast to SNe Ia or angular/angle-averaged BAO.

Observed values of $H(z)$ can be used to estimate the free parameters of the model and also the best-value for $H_0$ by minimizing the quantity
\begin{equation}
\chi^2_H (H_0, \theta_i)= \sum^{18}_{j=1} \frac{\left[H(z_j; \theta_i)-H_{obs}(z_j)\right] ^2}{\sigma^2_H(z_j)},
\end{equation}
where $\sigma^2_H$ are the measurement variances. $H_0$ has been fixed from Ref. \cite{Riess}, $H_{0} = 73.8\pm2.4$. The vector of model parameters, $\theta_i$, in our case will be $\theta_i= (\Omega_m, \nu)$. 

Noteworthy that constraining the free parameters of this model via $H(z)$ is controversial because
this is not a direct comparison with the ZKDR luminosity distance, but we have calculated $H(z)$ from the expression

\begin{equation}
H(z)=\left({\frac{d}{dz}\left({\frac{d_L}{(1+z)}}\right)}\right)^{-1},
\end{equation}
connecting in this way $H(z)$ with the parameter $\nu$ (or $\alpha$). We note that we are mixing the homogeneous FRW $H(z)$ with the inhomogeneous proposal, however a joint analysis with other probes makes sense and has been addressed in previous works \cite{Busti1}.

\section{Results}

The results of our analysis are shown in Tables \ref{T:1}-\ref{T:7}. We made the analysis with the parameter $\nu$, but in several tables are included the results for both parameters, $\nu$ used by Kantowski, and $\alpha$ related with $\nu$ by Eq.(\ref{alpha-nu}).

\begin{figure}
\includegraphics[width=0.45\textwidth]{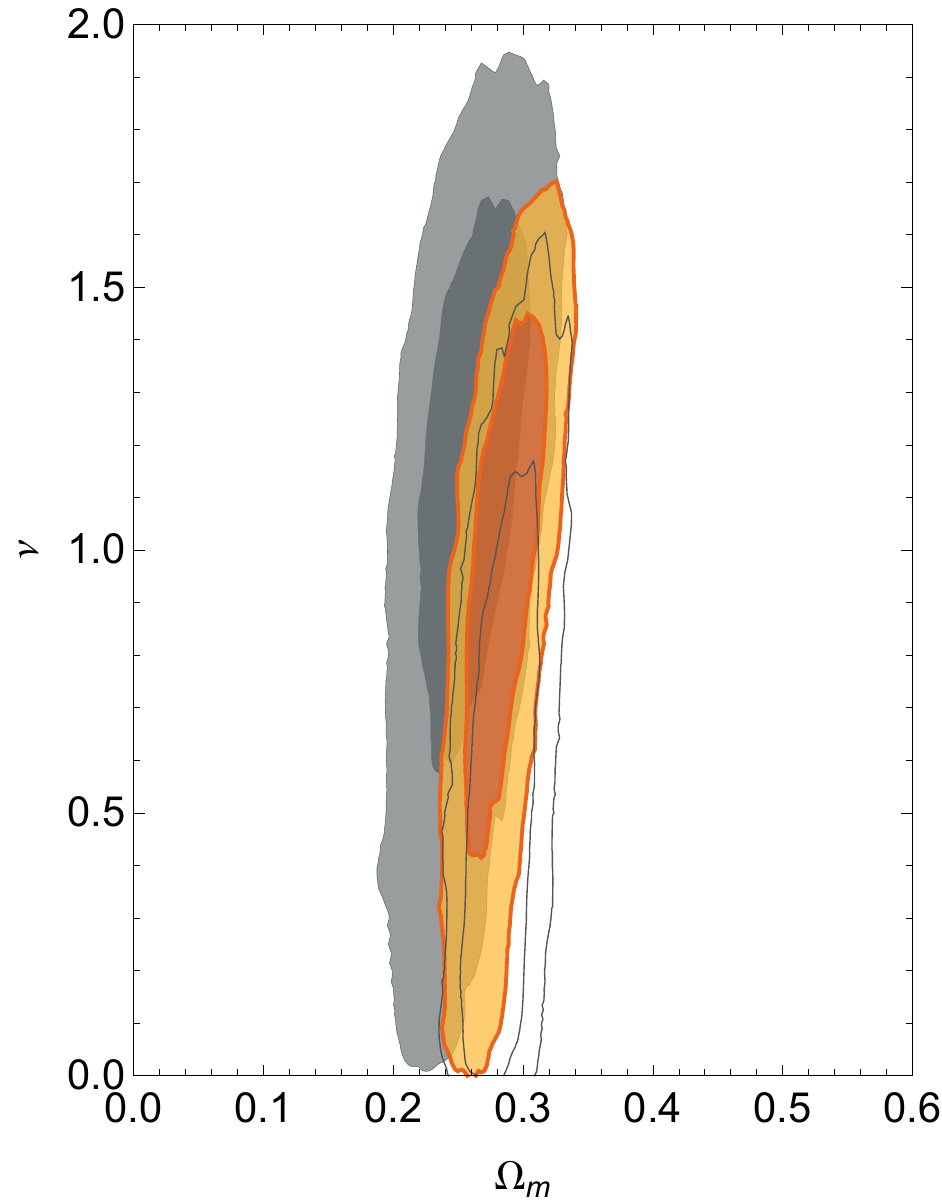}
\caption{Confidence regions in the ($\Omega_m$, $\nu$) plane for the model with a ZKDR luminosity distance using a prior on $\Omega_m$. The contours correspond to 1$\sigma$-2$\sigma$ confidence regions using: LGRBs, largest region on the back; SNe Ia, smallest region on the front; the combination of the two observational data, the region between the LGRBs region and SNe Ia region.}
\label{Fig:Contours}
\end{figure}

In Tables \ref{T:1} and \ref{T:2} are displayed the best fits for $\Omega_m$, $\nu$ (or $\alpha$) from the SNe Ia and GRBs probes reported in Ref. \cite{Union21} and Ref. \cite{Yonetoku12}, respectively. 
Table \ref{T:1} corresponds to the analysis assuming a Gaussian prior on $\Omega_m=0.266 \pm 0.029$ from WMAP-7 years \cite{WMAP}. Fig. \ref{Fig:Contours} shows the corresponding contour levels in the ($\Omega_m$, $\nu$) plane.

Table \ref{T:2} is the same as Table \ref{T:1} but without assuming any prior. The inclusion of a prior does not have a dramatic effect, but there is an improvement in the GRBs analysis, that goes from a $\chi^2_{\rm red}= 0.84$ without prior to $\chi^2_{\rm red}= 0.877 $ when the prior is assumed. The values of the smoothness parameter show a degree of clumpiness in the cosmic fluid, however it does not have an impact on reducing the proportion of dark energy, that remains over the 70\% of the total density; more precisely $\Omega_{\Lambda}=1- \Omega_{m}=0.716^{+0.020}_{-0.021}$ from the joint analysis of SNe Ia and GRBs. 
These results make us think that the DR approach does not describe the effects of clumpiness in a reliable way.

\begin{table}
\begin{ruledtabular}
\begin{tabular}{ccccc}
Sample & $\Omega_m$               & $\nu$   & $\alpha$ & $\chi^2_{\rm red}$  \\ \hline
SNe Ia   & $0.285^{+0.019}_{-0.018}$& $0.555^{+0.417}_{-0.363}$& $0.856^{+0.106}_{-0.176}$  & 0.975\\
GRBs   & $0.259^{+0.028}_{-0.028}$& $1.152^{+0.332}_{-0.421}$& $0.587^{+0.201}_{-0.202}$  & 0.877 \\
Joint & $0.284^{+0.021}_{-0.020}$& $0.963^{+0.316}_{-0.387}$& $0.685^{+0.164}_{-0.171}$  & 0.975 \\
\end{tabular}
\caption{Summary of the best estimates of model parameters ($\Omega_m$, $\nu$), obtained from the ZKDR luminosity distance using a prior on $\Omega_m$. The respective samples are SNe Ia reported in Ref. \cite{Union21} and GRBs reported in Ref. \cite{Yonetoku12}. The errors are at 68.3$\%$ confidence level. Joint stands for the joint analysis SNe Ia + GRBs. The corresponding confidence regions are shown in Figure \ref{Fig:Contours}.}
\label{T:1}
\end{ruledtabular}
\end{table}
\begin{table}
\begin{ruledtabular}
\begin{tabular}{ccccc}
Sample     & $\Omega_m$   & $\nu$    & $\alpha$& $\chi^2_{red}$  \\ \hline
SNe Ia     & $0.302^{+0.032}_{-0.026}$& $0.770^{+0.504}_{-0.469}$& $0.773^{+0.162}_{-0.256}$  & 0.975\\
GRBs       & $0.234^{+0.098}_{-0.055}$& $0.979^{+0.635}_{-0.641}$& $0.677^{+0.248}_{-0.380}$  & 0.840 \\
Joint     & $0.310^{+0.037}_{-0.033}$& $1.220^{+0.336}_{-0.444}$& $0.548^{+0.212}_{-0.222}$  & 0.973 \\
\end{tabular}
\caption{Summary of the best estimates of model parameter ($\Omega_m$, $\nu$). Same samples as in Table \ref{T:1} but, in this case, no prior in $\Omega_m$ is assumed.  The errors are at 68.3$\%$ confidence level.}
\label{T:2}
\end{ruledtabular}
\end{table}

Tables \ref{T:3} and \ref{T:4} correspond to the $2 \sigma$ confidence level constraints of the cosmological 
parameters, $\Omega_m$ and $\alpha$ with and without the prior on $\Omega_m$, respectively. 

Table \ref{T:4}, showing the 2-$\sigma$ confidence level for $\Omega_{m}$ and $\alpha$ without assuming a prior on $\Omega_m$, can be directly compared with the results given in Ref. \cite{Busti2} (that we include for completeness in table \ref{T:5}). In Ref. \cite{Busti2} it is confronted the ZKDR luminosity distance versus SNe Ia and GRBs probes with the samples data 557 SNe Ia Union2 \cite{Amanullah} and the 59 Hymnium GRBs \cite{Wei2010}. Using a $\chi^2$ minimization, their results, shown in Table \ref{T:5}, did not constrain the smoothness parameter.
The best fits obtained by combining the SNe Ia and GRBs probes are $\Omega_m=0.27$ and $\alpha=1$, with a $\chi^2=568.36$ or $\chi^2_{\rm red}=0.927$.

\begin{table}
\begin{ruledtabular}
\begin{tabular}{cccc}
Sample  & $\Omega_m (2\sigma)$   & $\alpha (2\sigma)$  & $\chi^2_{red}$  \\ \hline

SNe Ia                  & $0.250 \le \Omega_m \le 0.325 $& $0.461 \le \alpha \le 0.995 $& 0.975\\

GRBs                   & $0.204 \le \Omega_m \le 0.315 $& $0.183 \le \alpha \le 0.952 $ & 0.877 \\

Joint &  $0.246 \le \Omega_m \le 0.327 $ & $0.344 \le \alpha \le 0.965 $& 0.975
\end{tabular}
\caption{Limits to $\Omega_m$ and $\alpha$ with prior on $\Omega_m$.  The errors are at 95$\%$ confidence level. Joint means the combined analysis with SNe Ia + GRBs.}
\label{T:3}
\end{ruledtabular}
\end{table}
\begin{table}
\begin{ruledtabular}
\begin{tabular}{ccccc}
Sample & $\Omega_m (2\sigma)$   & $\alpha (2\sigma)$    & $\chi^2_{red}$  \\ \hline
SNe Ia  & $0.252 \le \Omega_m \le 0.373 $& $0.247 \le \alpha \le 0.991 $& 0.975\\
GRBs    & $0.142 \le \Omega_m \le 0.447 $& $0.072 \le \alpha \le 0.988 $ & 0.840 \\
Joint &  $0.248 \le \Omega_m \le 0.385 $ & $0.150 \le \alpha \le 0.940 $& 0.973
\end{tabular}
\caption{Limits to $\Omega_m$ and $\alpha$ without prior on $\Omega_m$.  The errors are at 95$\%$ confidence level.}
\label{T:4}
\end{ruledtabular}
\end{table} 
\begin{table}
\begin{ruledtabular}
\begin{tabular}{ccccc}
Sample &$\Omega_m (2 \sigma)$ & $\alpha (2 \sigma)$ & $\nu (2 \sigma)$ & $\chi^2_{\rm red}$\\ \hline
SNIa& $0.24 \le \Omega_m \le 0.35 $&$0.25 \le \alpha \le 1 $ &$0 \le \nu \le 1.68 $& 0.984\\
GRBs& $0.11 \le \Omega_m \le 1 $&  unconstrained &unconstrained& 0.411 \\
Joint&$0.24 \le \Omega_m \le 0.33 $ &$0.52 \le \alpha \le 1 $ &  $0 \le \nu \le 1.27 $& 0.927\\
\end{tabular}
\caption{Obtained constraints for $(\Omega_m, \alpha)$ in Ref. \cite{Busti2}; SNe Ia data from Ref. \cite{Amanullah} and GRBs data taken from Ref. \cite{Wei2010}. The smoothness parameter remains unconstrained from GRBs.}
\label{T:5}
\end{ruledtabular}
\end{table}

Particularly, for GRBs, using a better sample and the MCMC method, we obtained a reliable constraint with a $\chi^2_{\rm red}= 0.877$ from GRBs only, and $\chi^2_{\rm red}= 0.973$ from the joint analysis without assuming the prior on $\Omega_m$ and $\chi^2_{\rm red}= 0.975$ from the joint analysis assuming the prior on $\Omega_m$.

In a previous work done by Busti et al. \cite{Busti1}, a chi-square analysis was performed with Hubble parameter measurements reported in Ref. \cite{Simon2005} and in Ref. \cite{Daly2008}. Our analysis provides better constraints, that are summarized in Table \ref{T:6a}; while in Table \ref{T:6} we make the comparison between the results obtained in Ref. \cite{Busti1} and ours.

\begin{table}
\begin{ruledtabular}
\begin{tabular}{ccccc}
      & $\Omega_m (2\sigma)$ & $\alpha (2\sigma)$ & $\chi^2_{red}$  \\ \hline
No Prior & $0.208 \le \Omega_m \le 0.376 $& $0.510 \le \alpha \le 0.995 $& 1.032\\
With Prior & $0.223 \le \Omega_m \le 0.315 $& $0.626 \le \alpha \le 0.996 $ & 1.025 
\end{tabular}
\caption{Limits to $\Omega_m$ and $\alpha$ from $H(z)$. The second row is obtained assuming a prior of
$\Omega_m=0.266 \pm 0.029$ from WMAP-7 years \cite{WMAP}. The errors are at 95$\%$ confidence level.}
\label{T:6a}
\end{ruledtabular}
\end{table}

\begin{table}
\begin{ruledtabular}
\begin{tabular}{ccc}
Sample& $\Omega_m $  &  $\alpha$ \\ \hline
182 SNe Ia +Hubble&  $0.27 \le \Omega_m \le 0.37$& $0.66 \le \alpha \le 1$   \\
12 Hubble &  $0.27 \le \Omega_m \le 0.3$& $0.5 \le \alpha \le 1$   \\        
\end{tabular}
\caption{Summary of the best estimates of model parameters for the Hubble probe comparing the previous results by Busti (2011) \cite{Busti1} from samples reported in Ref. \cite{Simon2005} and Ref. \cite{Daly2008} and in the second row the ones presented in here only from the Hubble parameter given in the sample by Moresco (2012) \cite{Jimenez12}}
\label{T:6}
\end{ruledtabular}
\end{table}
Considering that the samples  of SNe Ia and GRBs correspond to disjoint intervals of redshift, the different values of $\alpha$ for each probe can be interpreted as a dependence of the clumpiness parameter  $\alpha$ on the redshift (see Table \ref{T:7}).
\begin{table}
\begin{ruledtabular}
\begin{tabular}{ccccc}
Sample& $ \Omega_m$ & $\alpha $ &  redshift range & $\chi_{\rm red}^2$ \\ \hline
SNe Ia & $0.285^{+0.019}_{-0.018}$ & $0.856^{+0.106}_{-0.176}$ &$0.015 \le  z \le 1.414$& 0.975   \\
GRBs &  $0.259^{+0.028}_{-0.028}$ & $0.587^{+0.201}_{-0.202}$ &$1.547 \le z \le 3.57$&0.877\\ 
Hubble &  $0.268^{+0.023}_{-0.023}$& $0.895^{+0.076}_{-0.122}$ & $0.09 \le z \le 1.75$ & 1.025  \\        
 Joint & $0.275^{+0.019}_{-0.018}$& $0.821^{+0.110}_{-0.129}$& $0.015 \le z \le 3.57$ &0.974 \\
\end{tabular}
\caption{Summary of the best estimates of model parameters and the corresponding redshift range using in all the cases a prior on $\Omega_m$ from Ref. \cite{WMAP}.
The smoothness parameter $\alpha$ shows a dependence on the redshift range. Joint: SNe Ia + Hubble + GRB.}
\label{T:7}
\end{ruledtabular}
\end{table}

\section{Conclusions}

We probed the ZKDR luminosity distance proposed by Dyer and Roeder (DR), that includes the effect of local inhomogeneities through the smoothness parameter $\alpha$, with data from SNe Ia \cite{Union21} and GRBs \cite{Yonetoku12}. Additionaly we tested $d_L$ with the direct Hubble measurements reported in Ref. \cite{Jimenez12}.

Better data and a refined statistics for GRBs allow us to probe the ZKDR luminosty distance in a range of higher redshifts with confident results. Indeed, our analysis improves, in great deal, previous works that did not succeed in constraining the smoothness parameter. 
 
We can summarize the presented results as follows:

1) We have probed in a reliable way two distinct redshift ranges, in fact two disjoint intervals of $z$:
the SNe Ia range, $0.015 \le z \le 1.414$ and $1.547 \le z \le 3.57$ for GRBs.

The obtained results, two different values of $\alpha$ in each interval, strongly suggest a dependence of the clumpiness parameter on the redshift, see Table \ref{T:7} for the values of $\alpha$ and the corresponding redshift range. Since the redshift intervals are disjoint, the different values of $\alpha$ show a kind of evolution in the smoothness parameter with respect to the redshift. In principle, assuming that this approach describes the clumpiness effects with accuracy, one would expect that nearby regions should correspond to more clumpiness, then to smaller values of $\alpha$. In other words, the structure formation process leads to a more locally inhomogeneous universe and thus, the smoothness parameter should evolve from homogeneity ($\alpha =1$) to  total clumpiness ($\alpha=0$). However, the test turns out on the contrary: the smoothness parameter evolves toward homogeneity, being $\alpha = 0.587^{+0.201}_{-0.202}$ for the range $1.547 \le z \le 3.57$ and a greater value, $\alpha= 0.856^{+0.106}_{-0.176}$ for the SNe Ia range, $0.015 \le z \le 1.414$. These results show that the Dyer-Roeder equation for $d_L$ should be revised, as several authors have suggested \cite{Clarkson2012}.

2) Our results indicate the existence of clumpiness, because the values of $\alpha$ differ from the homogeneous FRW model ($\alpha=0.856$, $\alpha=0.587$ and $\alpha=0.895$ from SNe Ia, GRBs and Hubble, respectively), however this does not have a clear impact in the amount of dark energy needed to fit the observations. According to SNe Ia, Hubble and GRBs probes, the amounts of dark energy that fit to observations are: 
$\Omega_{\Lambda}=0.715^{+0.018}_{-0.019}$,  $\Omega_{\Lambda}=0.732 \pm 0.023$ and $\Omega_{\Lambda}=0.741 \pm 0.028$, respectively (with prior on $\Omega_m$); in other words, the average matter density increases in the range of small $z$ ($\Omega_m=0.285^{+0.019}_{-0.018}$, from SNe Ia compared with the value in the range of GRBs, $\Omega_m=0.259^{+0.028}_{-0.028}$).

In conclusion, our results indicate that DR approach models in an incorrect way the effect of local inhomogeneities 
in the propagation of light. In Section II we have noticed that the model is formulated with at least two questionable
assumptions: that the Hubble parameter and the affine parameter as a function of the redshift have the same expressions than in the homogeneous FRW model.

So far, the evidence indicates that a modification of the Dyer-Roeder equation is necessary to describe in a reliable way the backreaction effects due to local inhomogeneities. It is worth then to test with observational data some of the improved models that try to mend the drawbacks of the DR model; for instance in Ref. \cite{Bolejko2011} it is considered a correction in the analogous of the smoothness parameter coming from density fluctuations along the line of sight.

\begin{acknowledgments}
A. M. acknowledges financial support by CONACyT (Mexico) through a PhD scholarship; partial support from CONACyT-Mexico through project 166581 is acknowledged.
\end{acknowledgments}

\end{document}